\documentclass[aps,twocolumn,showpacs,floats,superscriptaddress]{revtex4}
\usepackage{graphicx}
\usepackage{amssymb,amsmath}
\begin{document}

\author{L. Pollet}
\affiliation{Physics Department, Harvard University, Cambridge-MA, 02138, USA}
\affiliation{Department of Physics, University of Massachusetts, Amherst, MA 01003, USA}
\affiliation{Theoretische Physik, ETH Zurich, 8093 Zurich, Switzerland}
\author{J. D. Picon}
\affiliation{Theoretische Physik, ETH Zurich, 8093 Zurich, Switzerland}
\affiliation{ Institute of Theoretical Physics,  {\'E}cole polytechnique f{\'e}d{\'e}rale de Lausanne, Switzerland}
\author{H.P. B\"uchler}
\affiliation{Institute of Theoretical Physics III, Universit\"at Stuttgart,  70550 Stuttgart, Germany}
\author{M. Troyer}
\affiliation{Theoretische Physik, ETH Zurich, 8093 Zurich, Switzerland}

\title{Supersolid phase with cold polar molecules on a triangular lattice}

\date{\today}
\begin{abstract}
We study a system of heteronuclear molecules on a triangular lattice and analyze the potential of this system for the experimental realization of a supersolid phase.
The ground state phase diagram contains superfluid, solid and supersolid phases. At finite temperatures and strong interactions there is an additional emulsion region, in contrast to similar models with short-range interactions. 
We derive the maximal critical temperature $T_c$ and the corresponding entropy $S/N = 0.04(1)$ for supersolidity and find feasible experimental conditions for its realization.
\end{abstract}



\maketitle


Long-range interactions are a key ingredient in many models of strongly correlated electronic systems and frustrated quantum magnets~\cite{Schollwoeck04}. 
Magnetic dipolar interactions often occur in materials science and compete with short-range ferromagnetic interactions, which leads to spatially modulated 
phases~\cite{DeBell00}. The influence of long-range interactions is currently also attracting a lot of interest in cold atomic and molecular gases: the first
signatures of long-range interactions have been observed for magnetic interactions in ${}^{52}$Cr \cite{Griesmaier05, Lahaye07}, 
while electric dipole and van der Waals interactions between Rydberg states give rise to intriguing collective phenomena \cite{Heidemann07}.
In addition, there are big experimental efforts towards the realization of quantum degenerate polar molecules \cite{Sage05,Ni08,Deiglmayr08}, where the permanent dipole moment of
the molecules gives rise to a strong and highly tunable electric dipole-dipole interaction \cite{Micheli06,Buechler07,Pupillo08,Menotti08a,Menotti08b}.

\begin{figure}[h]
\centerline{\includegraphics[ width=\columnwidth]{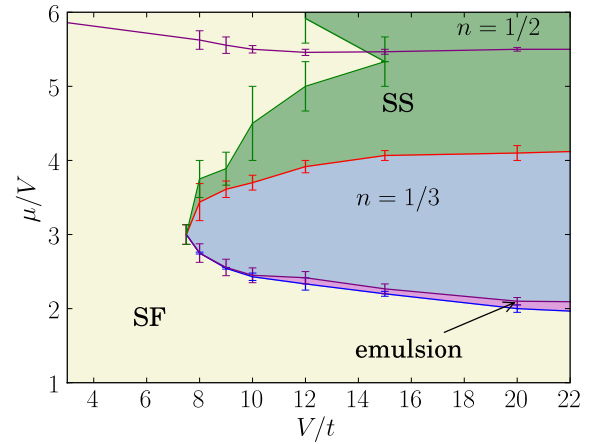}}
\caption{(Color online). 
Ground state phase diagram for the Hamiltionian Eq.(\ref{eq:Hamiltonian}) around $n=1/3$.  The phases are a superfluid `SF', supersolid `SS' and  a commensurate solid at density $n=1/3$. 
With the double line we indicate a transition region of the Spivak-Kivelson bubble type (emulsions) gradually going over to a region of incommensurate, floating solids with increasing interaction strength. For large interaction strength, and starting around half filling, the supersolid phase is suppressed by emerging solid ordering (stripes at half filling and incommensurate, floating solids at other fillings). }
\label{fig:phasediagram}
\end{figure}

In this Letter, we will concentrate on one intriguing aspect of dipolar systems, namely the possibility to observe a supersolid phase in a single component system. While the interpretation of super-phenomena observed with torsional oscillators in solid ${}^4$He remains a puzzle~\cite{Prokofev07}, supersolids might be much easier to realize in lattices. While there exist a number of lattice models of hard-core bosons \cite{Wessel05, Boninsegni05, Heidarian05, Melko05, Melko06, Hebert02, Schmid04, Batrouni95}, soft-core bosons \cite{Sengupta05} and quantum spins \cite{Ng06,Laflorencie07} most of these models are hard if not impossible to implement in a material.  Systems of cold hetero-nuclear molecules are described by similar Hamiltonians but with longer range dipolar interactions. Here, we demonstrate that they show supersolidity under feasible experimental conditions.

We consider bosonic polar molecules  in a strong electric field along the $z$-direction, which induces the dipole moment $d_{z} \leq d$; here $d$ denotes
the permanent dipole moment of the hetronuclear molecule. The dominant interaction between the polar molecules is then given by the dipole-dipole interaction
$V(R)= \frac{d_{z}^2}{4\pi \epsilon_0} \frac{R^2-3 z^2}{R^5}$, where the strength of the dipole interaction can be continuously tuned by the strength of the electric field.
In addition, the polar molecules are confined into the $xy$ plane by a strong transverse harmonic confinement as can be easily achieved by
a  strong one-dimensional standing laser  along the $z$ direction. The combination of strong transverse trapping and dipole interaction creates a repulsive barrier
\cite{Buechler07}, which prevents the collapse naturally present in bosonic dipolar gases \cite{Koch08}. The effective long-range two-dimensional potential is then found by integrating over the $z$ direction, and reduces to the effective $2D$ interaction  $V_{\rm eff}^{2 {\rm D}} \sim D / r^3$. We refer to Refs.~\cite{Buechler07, Pupillo08} for a detailed discussion on how such a potential can be tailored. Subjecting the molecules to a triangular lattice created by three lasers in the $xy$ plane,
a standard one-band tight binding analysis leads to an extended hard-core Bose-Hubbard model,
\begin{equation}
H = -t \sum_{\langle i,j \rangle } b_i^{\dagger} b_j + {\rm h.c.} + \frac{V}{2} \sum_{i\neq j} \frac{n_i n_j} { \vert {\bf R}_i - {\bf R}_j \vert^3} - \mu \sum_i n_i. \label{eq:Hamiltonian}
\end{equation}
Here, the first term describes the kinetic term with hopping amplitude $t$, $\mu$ is the chemical potential, $V=D/a^3$ the 2D effective potential amplitude ($a$ is the lattice spacing, and ${\bf R}_{i}$ denote the normalized lattice vectors). We will work with periodic boundary conditions, the hopping is set to unity, $t=1$, the linear system size is $L$ and the density denoted by $n = N/L^2$.

Our main results are the ground state phase diagram of Fig.~\ref{fig:phasediagram} featuring a superfluid, commensurate solid and supersolid phase and the finite temperature phase diagram for $V/t<15$ at constant density $n=0.4$. The entropy $S/N=0.04(1)$ of the supersolid phase at the highest $T_c$ is approximately one third of the entropy found in the system with short-range interactions at the same temperature and density, but is still comparable to the lowest entropies reached with bosonic ultracold alkali gases~\cite{Trotzky09}. We find transitions that are continuous or belong to the Spivak-Kivelson bubble-type transition~\cite{Spivak04}. 

We study the Hamiltonian Eq.~(\ref{eq:Hamiltonian}) by an unbiased and accurate quantum Monte Carlo simulation using the worm algorithm~\cite{Prokofev98} in the implementation of Ref.~\cite{Pollet07}.  To efficiently handle  the slow (but integrable) decay of the potential in two dimensions, we replace the potential by a tabulated potential which is summed over all periodic images (like in an Ewald summation) and which is only slightly different  from $1/r^3$. In contrast to the short-range model, simulations with repulsive long-range interactions are notoriously  more difficult and we suffer from the extra complication of a diverging number of low-energy metastable states~\cite{Menotti07}. Although this makes the identification of the phases and phase transitions at very large interactions $V$ hard for current computational techniques, the supersolid phase can still be  unambiguously  identified in the experimentally most relevant regime. To identify the superfluid, solid and supersolid phases we measure the superfluid density $\rho_s$ and the density wave structure factor $S_{\vec Q}/L^2 = \langle \vert \sum_{k=1}^{L^2} n_k e^{i {\vec Q} {\vec r_k}} \vert^2 \rangle / L^4$ with ${\vec Q}=(4\pi/3, 0)$.

\begin{figure}
\centerline{\includegraphics[width=\columnwidth]{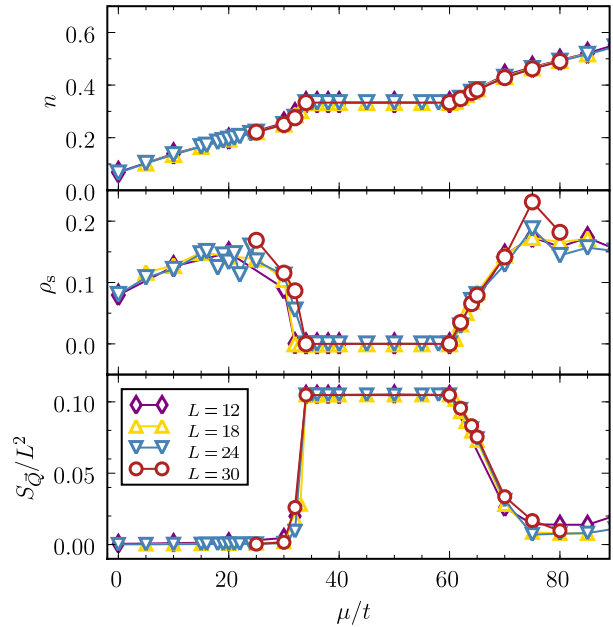}}
\caption{(Color online). Density $n$, superfluid density $\rho_s$ and structure factor $S_{\vec {Q}}/L^2$ for $V/t=15$ as a function of chemical potenial $\mu$. Statistical errors are smaller than symbol sizes if not shown.}
\label{fig:U15}
\end{figure}

\begin{figure}
\centerline{\includegraphics[width=\columnwidth]{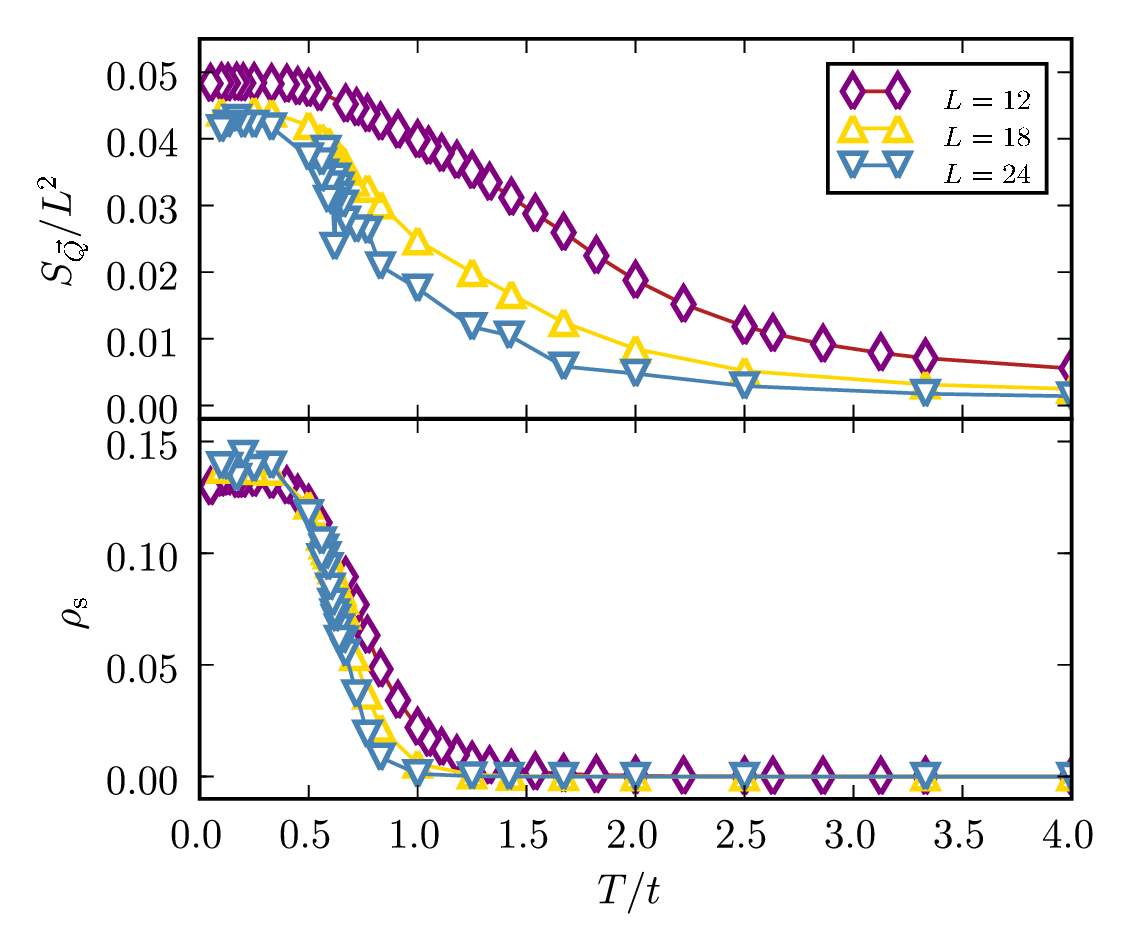}}
\caption{(Color online). Superfluid density $\rho_s$  and structure factor $S_{\vec{Q}}/L^2$  as a function of  temperature $T/t$ at fixed density $n=0.4$ and interaction strength $V/t=12$.  At low temperatures we have a supersolid phase with both commensurate, solid order and a finite superfluid density. Although small system sizes show a continuous transition reminiscent of a two-dimensional 3-state Potts transition, a careful analysis reveals non-monotonicity and hysteresis in the structure factor for larger system sizes, which is the first evidence for more complex physics in the emulsion phase. }

\label{fig:Temperature}
\end{figure}

\paragraph*{Ground state} We start our analysis with the ground state phase diagram, shown in Fig.~\ref{fig:phasediagram} for densities below half filling (the results above half filling are similar). This phase diagram has been obtained by extrapolating numerical data for different system sizes $L = 12, 18, 24$ and sometimes $L=30$ to infinity scaling the inverse temperature as $\beta t = L$. The quality of the raw data from which the phase diagram was obtained can be assessed in Fig.~\ref{fig:U15}.  For $V/t > 7.5(5)$ there is an insulating commensurate solid at density $n=1/3$. 

For densities below $1/3$ a superfluid phase is reached, similar to what is found for the short-range model~\cite{Wessel05, Boninsegni05, Heidarian05, Melko05, Melko06}, but the tranisition here is different and of  the  bubble type introduced by Spivak and Kivelson~\cite{Spivak04}: over a finite but narrow range of chemical potentials, small crystallites form an emulsion of bubbles inside a liquid. 

For $V=30$ (not shown) we find the first evidence for additional plateaus at various fillings below $n=1/3$ which are not present in the short-range model. With increasing system size the number of plateaus grows  and they are separated by small superfluid regions.  We expect that an incommensurate, floating solid is formed in the thermodynamic limit for strong interactions by analogy to the analysis of Ref.~\cite{Isakov07}.
 Note that in the classical limit of zero hopping the long-range model exhibits a devil's staircase (see Ref.~\cite{Bak82, Burnell09} for 1d) of various solid phases. 

Above the commensurate solid at $n=1/3$ we find  a continuous second-order phase transition belonging to the $3$D XY model universality class to a supersolid phase, similar to what occurs in the short-range model. While near the tip the supersolid phase exists only over a narrow density range, it quickly extends ($V/t=15$)  all the way to half filling. For larger interactions ($V/t > 20$) and close to half filling, the structure factor $S_{\vec Q}$ and the superfluid density go down and supersolidity is lost for $V/t=30$ at and near half filling. 

\begin{figure}
 \centerline{\includegraphics[width=\columnwidth]{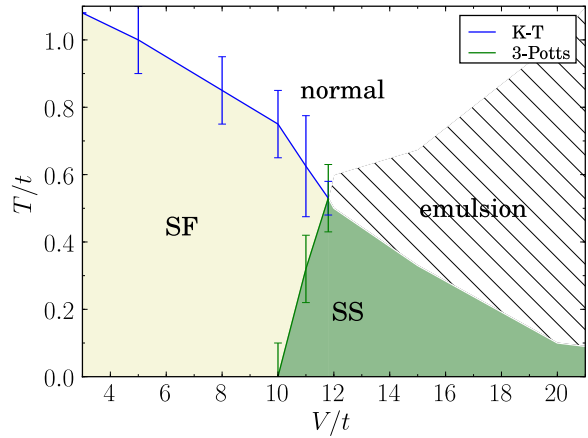}}
\caption{(Color online). Finite temperature phase diagram at fixed density $n=0.4$. The superfluid to normal liquid and the superfluid to supersolid transitions are found to be continuous (see text). After the Kosterlitz-Thouless (KT) and two-dimensional 3-state Potts transition lines cross, an intermediate phase of emulsions appears between the normal and the supersolid phases. The hatched region denotes where we found such emulsions for $L=24$. }
\label{fig:finitetemp_phasediagram}
\end{figure}

\paragraph*{Finite temperature} 
To study the  the transitions at finite temperature we will work in the canonical ensemble at a density $n=0.4$. An interaction strength of at least $V=10.0(5)$  is needed in order to observe a supersolid phase (see Fig.~\ref{fig:finitetemp_phasediagram}). For weak interactions we observe first a Kosterlitz-Thouless transition between a normal liquid and a superfluid phase, and a transition belonging to the two-dimensional 3-state Potts model then leads to the supersolid phase at lower temperature.
When those two continuous transition lines cross for $V/t=11.8(5)$ at a temperature $T_c=0.53(8)$,  the entropy per particle is $S/N=0.04(1)$ which is approximately one third of the entropy found in the short-range model with the same system parameters. For larger interaction strengths we observe in our simulations  the emergence of an emulsion region with many metastable states between the normal liquid and the supersolid phase. 
 More analysis for larger system sizes than what we can do in this study would be needed to accurately study the melting transition and the destruction of the superfluid order.

\paragraph*{Experimental proposal}   

We now outline optimal parameters for an experiment aiming at a homogeneous supersolid phase in thermodynamic equilibrium.
The experimentally optimal density for the observation of the supersolid phase is given by $n\approx0.4$, as it exhibits the largest superfluid fraction and consequently the highest critical temperature $T_c$. The supersolid phase exists over a finite range of densities in the phase diagram, and this allows for flatter curvature in the trap center than with parabolic traps.

From Fig.~\ref{fig:finitetemp_phasediagram} we see that the optimal interaction strength is $V/t= 11.8(5)$ with $T_c/t =0.53(8)$. The repulsion should not be much larger than $V/t=15$  because of the risk of hitting a large emulsion region from which experiments will be unable to equilibrate due to the existence of many metastable states. 
These parameters can be reached using {\rm LiCs} with a dipole moment $d=6.3 {\rm D}$ and an optical transition at $\lambda \approx 940 {\rm nm}$ for the
optical lattice \cite{Korek2000}. Then, a polarization of the molecules with $d_{z}\approx 0.1 d$ gives rise to the interaction energy $V \approx 0.23 E_{r}$ (here, $E_{r}\approx 1.6 {\rm kHz}$
denotes the recoil energy), and a relatively weak optical lattice with $V_{\rm \scriptscriptstyle lattice} \approx 8 E_{r}$ is sufficient to drive the system into the
supersolid phase; fine tuning of the parameters can be achieved by controlling the interaction strength via the static electric field. Note, that only a weak polarization of the molecules is required, and consequently, the supersolid phase  can also be reached for polar molecules with weaker dipole moments such as ${\rm Rb Cs}$ and ${\rm Li Rb}$.

Finally, we note that standard time-of-flight images are an easy and direct tool to identify the different phases.  In the superfluid phase the algebraic decay of the Greens function yields a strong signal at $k=0$ and at all reciprocal lattice vectors in the interference pattern. In the supersolid phase  three times more peaks will show up since the unit cell contains three lattice sites. 

In conclusion, we have shown that fully polarized molecules loaded into a triangular optical lattice exhibit a supersolid phase. For weak interactions  we have found in the ground state a bubble transition~\cite{Spivak04} between a superfluid ($n<1/3$) and a commensurate solid $( n = 1/3)$ while a 3D XY transition leads to supersolid phase at higher densities when increasing the chemical potential. At finite temperature and fixed generic density (say $n=0.4$), there is a Kosterlitz-Thouless transition from a normal liquid to a superfluid, and a 3-state Potts model transition to a supersolid. For larger interaction strengths ($V/t \ge 12.0(5)$), there is a wide range of temperatures in which an intermediate phase of emulsions is found.
The required temperature for reaching supersolidity is feasible and the wide range in density allows for flexibility. We suggest an optimal value  of $V/t \approx 12$ and a density of $n=0.4$ when $T_c \approx 0.53(8)$ or $S/N=0.04(1)$. Experimentally, the difference between the normal liquid, commensurate solid and supersolid phases can be detected by time-of-flight images. Our proposal is a very promising candidate for observing a clean supersolid phase in optical lattices.

We thank the Swiss National Science Fund and the Aspen Center for Physics for financial support. Simulations were performed on the Brutus cluster at ETH Zurich and use was made of the ALPS libraries for the error evaluation~\cite{ALPS}.
We wish to thank M. Greiner, S. Isakov, D. Pekker, N. V. Prokof'ev and S. Sachdev for interesting discussions, and thank the authors of Ref.~\cite{Capogrosso09} for sharing and discussing their related results on the square lattice prior to publication.

\end{document}